\documentclass[english,floatfix,11pt,nofootinbib,prd]{revtex4}
\usepackage[T1]{fontenc}
\usepackage[latin1]{inputenc}
\usepackage{amsmath}
\usepackage{amssymb}
\usepackage{graphicx}
\usepackage{esint}
\usepackage{longtable}
\usepackage{dcolumn}
\usepackage{babel}

\begin{document}

\title{Einstein contra Aristotle: the sound from the heavens}
\author{J. C. S. Neves}
\email{nevesjcs@ime.unicamp.br}
\affiliation{Instituto de Matemática, Estatística e Computação Científica, Universidade
Estadual de Campinas \\
 CEP. 13083-859, Campinas, SP, Brazil}

\begin{abstract}
In \textit{On the Heavens} Aristotle criticizes the Pythagorean point of view which claims the existence of a cosmic music and a cosmic sound. According to the Pythagorean argument, there exists a cosmic music produced by stars and planets. These celestial bodies generate sound in its movements, and the music appears due to the cosmic harmony. For Aristotle, there
is no sound produced by celestial bodies. Then, there is no music as well. However, recently, LIGO (Laser Interferometer Gravitational-Waves Observatory) has detected the gravitational waves predicted by Einstein. In some sense, a sound originated from black holes has been heard. That is, Einstein or the General Relativity and LIGO appear to be with the Pythagoreanism and against the master of the Lyceum. 
\end{abstract}

\maketitle

\section{Introduction}

The gravitational waves were confirmed by LIGO (Laser Interferometer Gravitational-Waves Observatory) Scientific Collaboration \cite{LIGO}. Such as the Einstein's General Theory of Relativity indicated, the motion of massive bodies may produce detectable waves in the spacetime fabric. Despite the Einstein's later position on the reality of the phenomenon \cite{Saulson}, the detection of the gravitational waves is a remarkable accomplishment of the Einsteinian theory.    

Due to its wavelengths, the gravitational waves detected by LIGO offer us the possibility of hearing them. In some sense, the gravitational waves phenomenon is able to generate sound from massive bodies in motion---in this case two black holes colliding. The LIGO has produced sound from these wavelengths with determined features\footnote{The audio from LIGO: https://www.ligo.caltech.edu/video/ligo20160211v2}.

In this note, the possibility of generating sound from celestial bodies in motion is stressed. The Collaboration renews an ancient debate. As we will see,  Aristotle denied the Pythagorean argument on the reality of both sound and music produced by celestial bodies in motion. This assumption made part of the Pythagorean thought on the musicality of the entire world. The master of the Lyceum, Aristotle, refused this interpretation from the sensible point of view. However, the modern science with its empirical obligation shows the contrary.

\section{The Aristotle's cosmos }

In his masterpiece on Cosmology and Astronomy, \textit{On the Heavens},
Aristotle explains his cosmos view. The Earth is at the center of
the universe. Planets and stars move around our planet in his geocentric
model. The cosmos is finite and made of elements contraries and intermediaries:
earth and fire are contraries; water and air are intermediaries. There
exists still a fifth element, the aether, which is part of the celestial
world. Each element tends to the their natural place, i.e., in the
absence of an external force, the elements follow to their natural
places. The celestial bodies are perfect\textemdash{}move in circular
trajectories (the perfect motion)\textemdash{}, eternal and immutable.
On the contrary, in the sub-lunar world, there exists the linear motion,
an imperfect movement. Moreover, in this region of the cosmos, one
has the generation and corruption. 

The Aristotle's cosmos was dominant until the Renaissance. With Copernicus
and Kepler, the West had a revolution: the Earth is not the center
of the cosmos, and the circular motion is not the celestial motion.
Planets, for example, according to Kepler's law, have an elliptical
trajectory and move around the Sun. 

In \textit{On the Heavens}, book II, especially in the chapter nine,
Aristotle criticizes the Pythagorean doctrine of the cosmic music.
Aristotle says that\footnote{The corresponding identification of the Aristotle's quotations in Ref. \cite{Aristotle} is given by the standard edition established by Bekker in $19th$ century.} \textquotedblleft{}it is clear that the theory
that the movement of the stars produces a harmony, i.e. that the sounds
they make are concordant, in spite of the grace and originality with
which it has been stated, is nevertheless untrue\textquotedblright{}
($290b12$). First of all, according to the ancient belief, an object in motion produces sound. For Aristotle, the planets and stars do not emit sound because these celestial objects are at rest in relation to the spheres or \textquotedblleft{}circles to which they are attached\textquotedblright{} ($289b30$). The apparent movement of these objects occurs because each one is fixed in a determined celestial sphere, and the spheres are not at rest. Moreover, he points out the fact that we do not hear these sounds. Therefore, due to these arguments, Aristotle refuses the Pythagorean celestial sound. 

\section{LIGO and the cosmic sound}

In another seminal work, Einstein in 1916 \cite{Einstein} obtained
the wave equation that describes a small perturbation in the spacetime
fabric. From the weak field approximation
\begin{equation}
g_{\mu\nu}=\eta_{\mu\nu}+h_{\mu\nu},
\end{equation}
where the curved spacetime is written as the flat spacetime $(\eta_{\mu\nu})$ plus a small perturbation $(h_{\mu\nu})$, Einstein showed that the small perturbation obeys a wave equation. Einstein called these waves of gravitational waves. This small perturbation, for example, may be generated by a massive  body in motion. 

Recently, the LIGO has made a tremendous announcement. In addition to the gravitational waves detection, the Collaboration made the first direct detection of black holes. The gravitational waves detected by LIGO were generated
by the motion, collision and merger of two black holes at a luminosity
distance of $410_{-180}^{+160}$ mega-parsecs. Their initial masses
were $36_{-4}^{+5}$ and $29_{-4}^{+4}$ solar masses. The final black
hole mass is $62_{-4}^{+4}$ solar masses. According to the energy
conservation, $3.0_{-0.5}^{+0.5}c^{2}$ solar masses were radiated
in gravitational waves ($c$ is the speed of light in vacuum). The
signal, the waves, had the frequency rangy from $35$ to $250$ Hertz.
The data was well described by General Relativity, and this announcement
confirmed the Einstein's theory once again.

During the announcement at press conference, González, a LIGO member, said with enthusiasm: \textquotedblleft{}We can hear the gravitational waves, we can hear the universe!\textquotedblright{}\footnote{In this link, the video with the announcement is showed (at 18:44, the González quotation): https://www.youtube.com/watch?v=aEPIwEJmZyE%
} But is it possible? According to the above argument, the perturbation\textemdash{}the
ripples of the spacetime\textemdash{}is described by waves, which
obey a determined wave equation. Any wave has a frequency. The waves
detected by LIGO, as we mentioned above, have a determined frequency
range. And within this range, it is possible a human being to hear. Then, using these frequencies which characterize these spectacular cosmic events, the LIGO Collaboration has created sound\textemdash{}which is a perturbation propagated
in a medium such as the air\textemdash{}and heard the cosmos in a sense. 

\textquotedblleft{}Some thinkers suppose that the motion of bodies
of that size must produce a noise\textquotedblright{}, said Aristotle
on the celestial bodies in $290b16$. But, he concludes,
\textquotedblleft{}melodious and poetical as the theory is, it cannot
be a true account of the facts\textquotedblright{} ($290b31$). However, contrary to the master of the Lyceum and his thought, massive bodies in motion may be heard.

\section{Conclusions}

One the most important predictions in General Relativity was confirmed: the gravitational waves were observed by LIGO Scientific Collaboration. And this remarkable result brings an ancient discussion to our modern view. The possibility of hearing  celestial bodies during its movements was an important position in Pythagorean philosophy, according to Aristotle. The master of the Lyceum denied this possibility. However, the LIGO experiment has produced sound from the wavelengths measured by the gravitational waves phenomenon. In this sense, Einstein or General Relativity and LIGO point out in the same direction of the Pythagorean school and against Aristotle.  

\begin{acknowledgments}
I would like to thank FAPESP (Fundação de Amparo à Pesquisa do Estado
de São Paulo) for the financial support (Grant No. 2013/03798-3) and
Elena Senik for reading the manuscript. 
\end{acknowledgments}

\end{document}